\documentclass{aa} 

\usepackage{graphicx} 
\usepackage{txfonts} 
\usepackage{natbib} 
 
\newcommand{\Mpc}{$h^{-1}$\thinspace Mpc} 
\newcommand{\Gpc}{$h^{-1}$\thinspace Gpc}

\def\apj{ApJ} 
\def\apjl{ApJL} 
\def\apjs{ApJS} 
\def\aj{AJ} 
\def\aap{A\&A} 
\def\mnras{MNRAS}

\begin{document}    
 
\title{Evolution of superclusters in the cosmic web } 
 
\author{J. Einasto\inst{1,2,3} 
\and  I. Suhhonenko\inst{1} 
\and  L. J. Liivam\"agi\inst{1} 
 \and M. Einasto\inst{1}   
}
\institute{Tartu Observatory, EE-61602 T\~oravere, Estonia 
\and  
ICRANet, Piazza della Repubblica 10, 65122 Pescara, Italy 
\and 
Estonian Academy of Sciences, 10130 Tallinn, Estonia
} 
 
\date{ Received 17 October 2018/ Accepted 26 January 2019}  
 
\authorrunning{Einasto et al.} 
 
\titlerunning{Evolution of Superclusters} 
 
\offprints{J. Einasto, e-mail: jaan.einasto@to.ee} 
 
\abstract {} {We investigate how properties of the ensemble of
  superclusters in the cosmic web evolve with time.}
{We perform numerical simulations of the evolution of the cosmic web
  using the $\Lambda$CDM model in box sizes $L_0=1024,~512,~256$~\Mpc.
  We find supercluster ensembles of models for four evolutionary
  stages, corresponding to the present epoch $z=0$, and to redshifts
  $z=1$, $z=3$, and $z=10$. We calculate fitness diameters of
  superclusters defined from volumes of superclusters divided to
  filling factors of over-density regions.  Geometrical and fitness
  diameters of largest superclusters, and the number of superclusters
  as functions of the threshold density are used as percolation
  functions to describe geometrical properties of the ensemble of
  superclusters in the cosmic web.  We calculate distributions of
  geometrical and fitness diameters and luminosities of superclusters,
  and follow time evolution of percolation functions and supercluster
  distributions. We compare percolation functions and supercluster
  distributions of models and samples of galaxies of the Sloan Digital
  Sky Survey (SDSS).  }
{Our analysis shows that fitness diameters of superclusters have a
  minimum at certain threshold density.  Fitness diameters around
  minima almost do not change with time in co-moving coordinates.
  Numbers of superclusters have maxima which are approximately
  constant for all evolutionary epochs. Geometrical diameters of
  superclusters decrease during the evolution of the cosmic web;
  luminosities of superclusters increase during the
  evolution.  }
{Our study suggests that evolutionary changes occur inside dynamical
  volumes of superclusters. The stability of fitness diameters and
  numbers of superclusters during the evolution is an important
  property of the cosmic web. }


\maketitle

\section{Introduction}

The large-scale distribution of galaxies in the Universe is 
very complex.  There exist  density enhancements 
of different size and shape, such as clusters of galaxies, filaments, walls, and
low-density regions (voids) between  high-density regions.
Largest building blocks of the Universe are superclusters of galaxies.
The supercluster concept was introduced by
\citet{de-Vaucouleurs:1953aa, de-Vaucouleurs:1958ab} for the Local or
Virgo supercluster.  Superclusters as clusters of rich clusters of
galaxies were defined by \citet{Abell:1958bs,Abell:1989fy}.  Actually
superclusters are much richer; they contain, in addition to rich Abell
type clusters, poor \cite{Zwicky:1968} clusters and galaxies. But most
importantly, cluster and galaxy filaments link superclusters to a
connected network, called cellular structure \citep{Joeveer:1978pb},
supercluster-void network \citep{Einasto:1980}, or cosmic web
\citep{Bond:1996fv}.

Cosmic web elements can be selected using various methods.
\citet{Cautun:2014qy} gives a good overview about various structure
finding algorithms.  Among these methods is the multiscale morphology
filter by \citet{Aragon-Calvo:2010wd}, Bayesian sampling of the
density field by \citet{Jasche:2010rt}, and many other methods. 
The largest elements of the cosmic web are superclusters of galaxies.
The definition of superclusters is not very precise since they have no
well-fixed boundaries.  Catalogues of rich clusters of galaxies by
\citet{Abell:1958bs,Abell:1989fy} were used by \citet{Einasto:1994wd,
  Einasto:1997lh, Einasto:2001oq} to compile all-sky catalogues of
superclusters.  The luminosity density field method was used by
\citet{Einasto:2007tg}, based on Two degree Field (2dF) redshift
survey. \citet{Costa-Duarte:2011ys}, \citet{Luparello:2011fr} and
\citet{Liivamagi:2012} used the Sloan Digital Sky Survey (SDSS) for
supercluster search.  { \citet{Chon:2015fk} analysed the definition of
superclusters and suggested to use the term ``superstes-clusters'' for
overdense regions which would eventually collapse in the future. }

To identify structures in the density field, it is necessary to define
a density threshold to separate high-density regions (superclusters)
from low-density regions (voids).  There is no natural value of the
threshold density. \citet{Costa-Duarte:2011ys} applied for the
selection of superclusters two criteria, one threshold density which
maximizes the number of superclusters, and the other which selects the
largest supercluster length (diameter) $\approx 120$~\Mpc, as adopted  by
\citet{Einasto:2007tg}.   
\citet{Liivamagi:2012} used for supercluster search two methods, one
with a fixed density threshold, and the other with an adaptive
density threshold, depending on the distribution of galaxies in the
particular region.

Large-scale systems of galaxies remember their
history well since the crossing time in these systems is much greater
than in small systems \citep{Joeveer:1977lj}.  The evolution of the
cosmic web can be investigated by numerical simulations, and results
of simulations can be compared with observations.  These
studies have a long history \citep{Aarseth:1979, Doroshkevich:1982fk,
  Zeldovich:1982kl, White:1983}. Recent advances in the study of the
cosmic web and its evolution are summarised in the Zeldovich Symposium
report \citep{VandeWeygaert:2016zt}.  In most studies the evolution of
the whole web is considered.  Special studies are devoted to investigate
the evolution of components of the web, such as clusters and voids.
\citet{Luparello:2011fr} and \citet{Gramann:2015gf} investigated the
future evolution of superclusters as virialised structures.

The goal of the present study is to investigate the evolution of the
ensemble of superclusters in the cosmic web.  Superclusters are the
largest known coherent structures of the Universe.  In the formation
of superclusters large-scale density perturbations play an important
role. To include large-scale density perturbations we performed
numerical simulations of the evolution in a box of size
$L_0=1024$~\Mpc.  As shown by \citet{Klypin:2018aa}, larger simulation
boxes are not needed to understand main properties of the cosmic web.
For comparison we also used simulations in smaller boxes of sizes
$L_0=512,~256$~\Mpc.

To describe geometrical properties of the ensemble of superclusters in
the cosmic web we shall use the extended percolation analysis by
\citet{Einasto:2018aa}.   A critical parameter in the search of
superclusters is the density threshold to divide the density field
into high- and low-density regions. In percolation analysis
high-density regions are called clusters, and low-density regions
voids \citep{Stauffer:1979aa}.  We use density fields smoothed with
8~\Mpc\ kernel.  In this case high-density regions can be called
superclusters.  We shall find ensembles of superclusters of models for
four epochs, corresponding to the present epoch $z=0$, and to
redshifts $z=1$, $z=3$, and $z=10$.  We vary the density threshold in
broad limits, divide the density field at each threshold density into
high- and low-density systems, and select the largest
superclusters. Lengths and volumes of largest superclusters, and
numbers of superclusters at respective threshold density level, are
used as percolation functions.

{ In addition to geometrical diameters of superclusters, we shall
  introduce in our analysis fitness volumes and diameters of
  superclusters. Fitness volumes are  
proportional to their geometrical volumes, weighted by a factor to get
for the sum of fitness volumes the whole volume of the sample.  We use
fitness volumes to calculate fitness diameters, and   use the
distribution of fitness diameters of largest superclusters as an
additional percolation function.}  Percolation functions are used to
describe properties of the whole ensemble of superclusters.  We also derive
distributions of sizes and masses of superclusters.  The comparison of
percolation functions and size and mass distributions for different
epochs allows to study the evolution of the ensemble of superclusters.
For comparison we use the main sample of the SDSS DR8 survey to
calculate the luminosity density field of galaxies, and to find
percolation functions of the SDSS sample.  Thorough this paper we use
the Hubble parameter $H_0 = 100 h$~km~s$^{-1}$~Mpc$^{-1}$.

The paper is organized as follows. In the next Section we describe the
calculation of the density field of observed and simulated samples,
the method to find superclusters and their parameters, and
supercluster fitness diameters. In Section 3 we perform percolation
analysis of simulated superclusters, and investigate changes of
percolation functions and supercluster parameters with time.  We also
compare percolation properties of model and SDSS samples, and the
dependence of percolation properties on parameters of the cosmic
model. The last Section brings the general discussion and summary
remarks.

\section{Data} 
 
To find superclusters we have to fix the supercluster definition
method and basic parameters of the method.  We shall use the density
field method. This method allows to use flux-limited galaxy samples,
and to take into account galaxies too faint to be included to the
flux-limited samples.  We define superclusters as large non-percolating
high-density regions of the cosmic web.  Based in our previous
experience we use for supercluster search the luminosity (matter in
simulations) density field, calculated with the $B_3$ spline of kernel
size $R_B=8$~\Mpc.  The determination of the second parameter of the
supercluster search, the threshold density, shall be discussed below.

\subsection{Simulation of the cosmic web }
 
We performed simulations in the conventional $\Lambda$CDM model with
parameters $\Omega_{\mathrm{m}} = 0.286$, $\Omega_{\Lambda} = 0.714$,
and $\Omega_{\mathrm{tot}} = 1.000$.  The initial density fluctuation
spectra were generated using the COSMICS code by
\citet{Bertschinger:1995}.  To generate the initial data we used the
baryonic matter density $\Omega_{\mathrm{b}}= 0.044$
(\citet{Tegmark:2004}). Calculations were performed with the GADGET-2
code by \citet{Springel:2005}.  Particle positions and velocities were
extracted for 7 epochs between redshifts $z = 30 \dots 0$. We shall
search for superclusters at four cosmological epochs, corresponding to
redshifts $z=0$, $z=1$, $z=3$ and $z=10$.  The resolution of all
simulations was $N_{\mathrm{part}}=N_{\mathrm{cells}}=512^3$, the size of the simulation
boxes was $L_0=1024$~\Mpc, the volume of simulation box was
$V_0=1024^3$~(\Mpc)$^3$, and the size of the simulation cell was
2~\Mpc.  This box size is sufficient to see the role of large-scale
density perturbations to the evolution of superclusters, which have
characteristic lengths up to $\sim 100$~\Mpc\
\citep{Liivamagi:2012}. We designate the simulation with the box size
$L_0=1024$~\Mpc\ as L1024.z, where the index $z$ notes the simulation
epoch  redshift.  To see the dependence of results on the size of the simulation
box we used also  simulations in $L_0=512$~\Mpc\ and $L_0=256$~\Mpc\
boxes; these simulations are designed as L512.z and L256.z.  Data on
simulated and SDSS superclusters are given in Table~\ref{Tab1}.

\subsection{SDSS data} 

The density field method allows to use flux-limited galaxy samples,
and to take statistically into account galaxies too faint to be
included to the flux-limited samples, as applied among others by
\citet{Einasto:2003a, Einasto:2007tg}, and \citet{Liivamagi:2012} to select
superclusters of galaxies. 

We use the Sloan Digital Sky Survey (SDSS) Data Release 8 (DR8)
\citep{Aihara:2011aa} and galaxy group catalogue by
\citet{Tempel:2012aa} to calculate the luminosity density field.  In
the calculation of the luminosity density field we need to take into
account the selection effects that are present in flux-limited samples
\citep{Tempel:2009sp, Tago:2010ij}. In the calculation of the
luminosity density field galaxies were selected within the apparent
{\em r} magnitude interval $12.5 \le m_r \le 17.77$
\citep{Liivamagi:2012}.  In the nearby region relatively faint
galaxies are included to the sample, in more distant regions only the
brightest galaxies are seen.  To take this into account, we calculate
a distance-dependent weight factor:
\begin{equation}
  W_L(d) =  {\frac{\int_0^\infty L\,\phi(L)\
      \mathrm{d}L}{\int_{L_1}^{L_2} L\,\phi(L)\,\mathrm{d}L}} ,
  \label{eq:weight}
\end{equation}
where $L_{1,2}=L_{\sun} 10^{0.4(M_{\sun}-M_{1,2})}$ are the luminosity
limits of the observational window at distance $d$, corresponding to
the absolute magnitude limits of the window $M_1$ and $M_2$.  The
weight factor $W_L(d)$ increases to $\approx 8$ at the far end of the
sample; for a more detailed description of the calculation of the
luminosity density field and corrections used see
\citet{Liivamagi:2012}.    The 
algorithm to find superclusters is described below.  The volume of the
SDSS main galaxy sample is (509~\Mpc)$^3$ \citep{Liivamagi:2012}.

\subsection{Calculation of the density field} 
 
We determined the density field using a $B_3$ spline
\citep[see][]{Martinez:2002fu}:
\begin{equation} 
B_3(x)=\frac1{12}\left[|x-2|^3-4|x-1|^3+6|x|^3-4|x+1|^3+|x+2|^3\right]. 
\end{equation} 
This function is different from zero only in the interval
$x\in[-2,2]$.  To calculate the high-resolution density field we use
the kernel of the scale, equal to the cell size of the simulation,
$L_0/N_{\mathrm{grid}}$, where $L_0$ is the size of the simulation box,
and $N_{\mathrm{grid}}$ is the number of grid elements in one
coordinate.  The smoothing with index $i$ has a smoothing radius
$r_i= L_0/N_{\mathrm{grid}} \times 2^i$. The effective scale of
smoothing is equal to $r_i$.  We applied this smoothing up to
index 6.  For models of the L1024 series smoothing index 2 corresponds
to the kernel of radius 8~\Mpc, for models of L512 and L256 series
smoothing indexes 3 and 4 correspond to kernel radius 8~\Mpc.  Most
calculations were performed with the model in the simulation box of
size $L_0=1024$~\Mpc, and with smoothing scale $R_B = 8$~\Mpc.  To see
the dependence of results on the smoothing scale we made calculations
for the $L_0=1024$~\Mpc\ model  using smoothing kernels of size
$R_B = 4$~\Mpc\ and $R_B = 16$~\Mpc.  These model series are noted as
F1024 for the $R_B = 4$~\Mpc\ case, and E1024 for the $R_B = 16$~\Mpc\
case { (F for Fine and E for Extended)}.

\subsection{Percolation functions and cluster parameters}
 
The percolation analysis consists of several steps: finding
over-density regions (clusters as potential superclusters) in the
density field, calculation of parameters of potential superclusters,
and finding the supercluster with the largest volume for a given density
threshold.   As traditional in the percolation
analysis, in general case over-density regions are called clusters
\citep{Stauffer:1979aa}.

We scan the density field in the range of threshold densities from
$D_t=0.1$ to $D_t=10$ in mean density units.  We use a linear step of
densities, $\Delta D_t = 0.1$, to find over- and under-density
regions. This range covers all densities of practical interest, since
in low-density regions the minimal density is $\approx 0.1$, and the
density threshold to find conventional superclusters is
$D_t \approx 5$ \citep{Liivamagi:2012}.  We mark all cells with
density values equal or above the threshold $D_t$ as filled
regions, and all cells below this threshold as empty regions.

Inside the first loop we make another loop over all filled cells
to find neighbours among filled cells. Two cells of the same type
are considered as neighbours (friends) and members of the cluster
if they have a common sidewall. Every cell can have at most six cells
as neighbours.  Members of clusters are selected using a
Friend-of-Friend (FoF) algorithm: the friend of my friend
is my friend.  To exclude very small systems, only systems having
fitness diameters at least 20~\Mpc\ are added to the 
list of over-density regions (see below for the definition of fitness
diameters).   

The next step is the calculation of parameters of clusters.  We
calculate the following parameters: centre coordinates,
$x_c, y_c, z_c$; diameters (lengths) of clusters along coordinate
axes, $\Delta x,~ \Delta y,~ \Delta z$; geometrical diameters
(lengths),
$L_{g} = \sqrt{(\Delta x)^2 + (\Delta y)^2 + (\Delta
    z)^2}$; fitness diameters (lengths), $L_f$, discussed in
the next subsection; geometrical volumes, $V_g$, defined as the volume
in space where the density is equal or greater than the threshold
density $D_t$; total masses (or luminosities), $\cal{L}$, the mass
(luminosity) inside the density contour $D_t$ of the cluster, in units
of the mean density of the sample.  We also calculate total volume of
over-density regions, equal to the sum of volumes of all clusters,
$V_C = \sum V_g$, and the respective total filling factor, 
\begin{equation}
F_f = N_f/N_{\mathrm{cells}}= V_C/V_0,
\label{fill}
\end{equation}
where  $N_f$ is the number of filled (over-density) simulation cells,
and  $V_0$ is the volume of the sample. 

During the cluster search we find the cluster with the largest volume
for the given threshold density. We store in a separate file for each
threshold density the number of clusters found, $N(D_t)$, and main data on
the largest cluster: the geometrical diameter, $L_{g}(D_t)$; the fitness
diameter, $L_f(D_t)$; the geometrical volume $V_g(D_t)$; the  mass
(luminosity) of the largest cluster, $\cal{L}(D_t)$, and the total
filling factor, $F_f(D_t)$.  Diameters are found in \Mpc,
volumes in cubic \Mpc,  masses/luminosities in units of the mean
density of the sample.  These parameters as functions of the density
threshold $D_t$ are called percolation functions.  They are needed to
characterise general geometrical properties of the ensemble of
superclusters in the cosmic web, and to select the proper threshold
density to compile the actual supercluster catalogue.  In total we
have for every evolutionary stage 100 catalogues of clusters
(over-density regions) as potential supercluster catalogues.
Notice that \citet{Einasto:2018aa} used  filling factor
of largest clusters, $\mathcal{F}(D_t) = V_{\mathrm{max}}/V_0$ as a
percolation function.
  
{ We calculated for each model the variance of the density contrast,
\begin{equation}
\sigma^2 = 1 /N_{\mathrm{cells}} \sum{(D(\mathbf{x})-1)^2},
\label{disp}
\end{equation}
where $D(\mathbf{x})$ is the density at location $\mathbf{x}$, and
summing is over all cells of the density field. The dispersion of the
density contrast $\sigma$ depends on the smoothing length $R_B$ and
the cosmic epoch $z$ of models, see below.}

{ In observational studies of superclusters, defined on the basis
  of luminosity density field, it is natural to use the density
  threshold in mean density units, $D_t$, to divide the field into
  high- and low-density regions.  We did all our calculations using
  density threshold in these units.  However, in theoretical
  interpretation of results it is more convenient to express densities
  and threshold densities in units of the dispersion of the density
  contrast \citep{Yess:1996aa, Sahni:1997ai, Colombi:2000fj}.  Thus we
  recalculated all percolation functions using as arguments density
  thresholds reduced to unite value of the dispersion of the density
  contrast:
\begin{equation}
  x= (D_t - 1)/\sigma.
  \label{sigma2}
\end{equation}
In the discussion below we use, depending on the task, threshold
densities in both units.
  }
 
\subsection{Supercluster fitness diameters}

We define the fitness volume of the supercluster, $V_f$, proportional
to its geometrical volume, $V_{g}$, divided by the total filling
factor:
\begin{equation}
V_f =V_{g}/F_f,
\end{equation}
or, using the definition of the total filling factor  of all over-density
regions at this threshold density, Eq.~(\ref{fill}),
\begin{equation}
V_f = V_g/V_C \times V_0.
\end{equation}
In this way we get for the sum of fitness volumes the volume of the
sample.  { In earlier percolation studies the volume (or the
  filling factor) of the largest cluster and the total filling factor
  were considered as separate characteristics \citep{Klypin:1993aa,
    Sahni:1997ai, Shandarin:1998aa}.  We combine these parameters into
  one new parameter. }  The fitness volume measures the ratio of the
supercluster volume to the volume of all superclusters (all filled
over-density regions) at the particular threshold density, multiplied
by the whole volume of the sample.  It has some analogy with the
fatness factor defined by \citet{Einasto:2018aa} as the ratio of the
volume of the cluster to its maximal possible volume for a given
geometrical diameter.  Fatness and fitness volumes of superclusters
measure the volume of the supercluster in different ways, in one case
in relation to its maximal possible value, and in the other case in
relation to the summed volume all superclusters.

Fitness diameters (lengths) of  superclusters are calculated from their fitness
volumes as follows: 
\begin{equation}
L_f = V_f^{1/3} = (V_g/V_C)^{1/3} \times L_0. 
\end{equation}
Fitness diameters of largest
superclusters are found for all threshold densities, $D_t$.  We use
fitness diameters of largest superclusters, $L_f(D_t)$, as
a percolation function, in addition to other percolation functions ---
geometrical diameters, $L_g(D_t)$, total filling factors, $F_f(D_t)$, and
numbers of clusters, $N(D_t)$.

At very small threshold densities the largest supercluster occupies
almost the whole volume of the samples.  Thus, by definition, the
fitness diameter at very small threshold densities is approximately
equal to the size of the sample, $L_f= L_0$.  At very high threshold
densities the largest supercluster is the only supercluster, its
volume is equal to the volume of all filled cells, and by definition
also $L_f = L_0$.  At medium threshold densities the volume of the
largest supercluster is smaller than the volume of all filled cells,
$V_g < V_C$, thus fitness diameters are smaller than the size of the
sample, and follow at threshold densities $D_t \le 2$ approximately
geometrical diameters.  However, geometrical diameters of largest
superclusters decrease with increasing threshold density almost
continuously.  In contrast, fitness diameters of largest superclusters
have a minimum at a certain threshold density.  This minimum shows
that the largest supercluster has the smallest volume fraction
$V_g/V_C$.  The minimum of the fitness diameter corresponds to the
maximum of the fragility of the supercluster as a function of
threshold density, and can be used as an additional parameter to
characterise the structure of the cosmic web at supercluster scales,
and to find the threshold density for supercluster selection.

\begin{figure*}[ht]
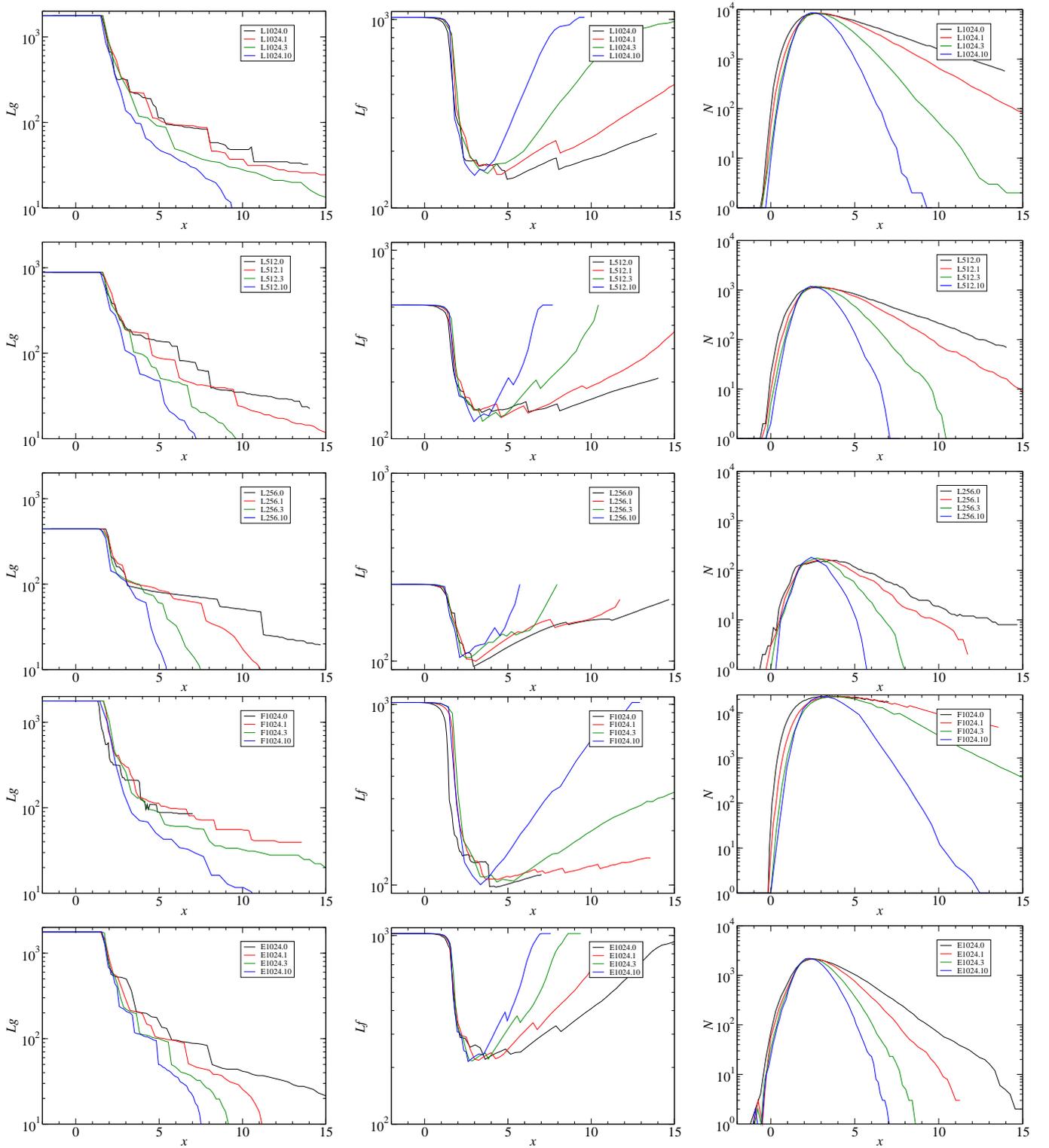
 
\centering 
\hspace{2mm}
\resizebox{0.31\textwidth}{!}{\includegraphics*{L1024_Lgeom_sigma.eps}}
\hspace{2mm} 
\resizebox{0.31\textwidth}{!}{\includegraphics*{L1024_Ldyn_sigma.eps}}
\hspace{2mm}
\resizebox{0.31\textwidth}{!}{\includegraphics*{L1024_N_sigma.eps}}\\
\hspace{2mm} 
\resizebox{0.31\textwidth}{!}{\includegraphics*{L512_Lgeom_sigma.eps}}
\hspace{2mm} 
\resizebox{0.31\textwidth}{!}{\includegraphics*{L512_Ldyn_sigma.eps}}
\hspace{2mm} 
\resizebox{0.31\textwidth}{!}{\includegraphics*{L512_N_sigma.eps}}\\
\hspace{2mm} 
\resizebox{0.31\textwidth}{!}{\includegraphics*{L256_Lgeom_sigma.eps}}
\hspace{2mm} 
\resizebox{0.31\textwidth}{!}{\includegraphics*{L256_Ldyn_sigma.eps}}
\hspace{2mm} 
\resizebox{0.31\textwidth}{!}{\includegraphics*{L256_N_sigma.eps}}\\
\hspace{2mm} 
\resizebox{0.31\textwidth}{!}{\includegraphics*{L1024_scl1_Lgeom_sigma.eps}}
\hspace{2mm}
\resizebox{0.31\textwidth}{!}{\includegraphics*{L1024_scl1_Ldyn_sigma.eps}}
\hspace{2mm} 
\resizebox{0.31\textwidth}{!}{\includegraphics*{L1024_scl1_N_sigma.eps}}\\
\hspace{2mm} 
\resizebox{0.31\textwidth}{!}{\includegraphics*{L1024_scl3_Lgeom_sigma.eps}}
\hspace{2mm}
\resizebox{0.31\textwidth}{!}{\includegraphics*{L1024_scl3_Ldyn_sigma.eps}}
\hspace{2mm} 
\resizebox{0.31\textwidth}{!}{\includegraphics*{L1024_scl3_N_sigma.eps}}\\
\hspace{2mm} 
\caption{{\em Left} panels show geometrical length functions; {\em
    middle} panels show  fitness length functions; {\em right}
  panels show number functions. As arguments of percolation functions
  we use the reduced threshold density, $x= (D_t - 1)/\sigma$. Panels
  from top down are for models L1024, ~L512, L256,~F1024, E1024.  }
\label{fig:evol} 
\end{figure*}

\section{Analysis of models}

\subsection{Percolation functions of L1024 model samples}

We use percolation functions to characterise geometrical properties of
the cosmic web and to select superclusters.  Superclusters are defined
as large non-percolating high-density regions of the density field,
smoothed with 8~\Mpc\ scale.  To select superclusters we have to find
proper value of threshold density to divide the density field to over-
and under-density regions. We shall use for this purpose percolation
functions.  { Fig.~\ref{fig:evol} shows geometrical length
  functions, $L_g$, fitness diameter functions, $L_f$, and numbers of
  clusters, $N$. Upper panels show these functions for the L1024
  model, in following panels for models of series L512, L256, F1024
  and E0124, all for redshifts $z=0$, $z=1$, $z=3$ and $z=10$.  In
  this Figure we use the reduced threshold density,
  $x= (D_t - 1)/\sigma$, as arguments of percolation functions.  }

{\scriptsize 
\begin{table*}[ht] 
\caption{Parameters of model and SDSS superclusters.} 
\centering 
\begin{tabular}{lllrrrrrrrrrrrc}
\hline  \hline
  Sample&$\sigma$ & $P$& $x_P$&$D_{\mathrm{max}}$ &$x_{\mathrm{max}}$&$N_{\mathrm{max}}$&
                $L_g$ & $L_f$ &$D_t$  &$x_t$ &$N_{scl}$ & $L_g$ & $L_f$ & $F_{f}$\\  
\hline  
(1)&(2)&(3)&(4)&(5) &(6)&(7)&(8)&(9)&(10)&(11)&(12)&(13)&(14)&(15)\\ 
\hline  \\
  L1024.0&0.6458 & 2.00 &1.6& 2.70 &2.6&8321  &  316 &178&
                                                       4.20&5.0 & 6044&  113  &142& 0.00788\\
  L1024.1&0.3683 & 1.60 &1.6& 2.10 &3.0&8472 &  317& 178 & 
                      2.70&4.6  &  6090 &   113  &  150 & 0.00760\\
  L1024.3&0.1852  & 1.30 &1.6& 1.50 &2.7&8535 &  348  & 190 &
                                                              1.70&3.8  &  6607 &   118 & 152 & 0.00930\\
  L1024.10&0.0667& 1.10 &1.5& 1.16&2.4&8643&   332  & 174 &
                                                            1.20&3.0  &  7833 &   137 & 149 & 0.01469\\
\\
  L512.0&0.6411 & 1.90 &1.4& 2.90  &3.0&  1120 &   244   & 152 & 3.60&
                                                                       4.1  & 995 &  156  &  140 & 0.01374\\
  L512.1&0.3703 & 1.50 &1.4& 2.10   &3.0& 1173  &  189  &  142 &
                                                                 2.70&4.6  &  835  &  95  & 130  & 0.00769\\
  L512.3&0.1869 &  1.30 &1.6& 1.50  &2.7&  1185 &   252 &  155 &
                                                                 1.65&3.5 &  1029  & 103 & 124 &0.01308\\
  L512.10&0.0676&1.10 &1.5& 1.16  &2.4&  1187 &   280 &   164 &
                                                                1.20&3.0 &  1072 &  108 &123 & 0.01528\\
\\
  L256.0&0.6129 & 2.00 &1.6& 3.30 &3.8&   158 &   90  &  102 &
                                                               2.80&2.9  &  147 &  134  &  94 & 0.02959\\
  L256.1&0.3582 & 1.50 &1.4& 2.10  &3.1&  169 &  107 &  100 &
                                                              2.10&3.1  &  169 &  107 &  100 & 0.02616\\
  L256.3&0.1823 &  1.25 &1.4& 1.50  &2.7&  178 &  122  & 104 &
                                                               1.45&2.5  &  164  & 133  & 104 & 0.04066\\             
  L256.10&0.0665&1.08 &1.2& 1.16&2.4&   183  &  135 &  109 &
                                                             1.14&2.1  &  159  & 143 &  104 & 0.05293\\
\\
  F1024.0&1.2829 & 2.70 &1.3& 5.30 &3.4&23819 &  211 &135&
                                                           4.80&3.0 & 23127 & 211& 133 & 0.01960\\ 
  F1024.1&0.6640 & 2.10 &1.7& 3.60 &3.9& 23309 & 133 &108&
                                                           3.50&3.8&  22663 &  134 &  107 &  0.01621\\
  F1024.3&0.2997 & 1.50 &1.7& 2.20 &4.0& 22680 & 126 &114 &
                                                            2.30&4.0& 22249 &  96&  104 & 0.00958\\
  F1024.10&0.1044& 1.15 &1.4& 1.35 &3.4& 22757 &   87& 100 &
                                                             1.35&3.4 & 22757 &  87&  100&  0.01254\\
\\
  E1024.0&0.3298 & 1.50 &1.5& 1.80  &2.4& 2139 &  526 &  285 &
                                                               2.20&3.6 & 1747  & 205 &  221 & 0.01622\\
  E1024.1&0.1998 & 1.30 &1.5& 1.50  &2.5& 2128 &  532 &  290 &
                                                               1.65&3.3  & 1911 &  214 &  218&  0.01927\\
  E1024.3&0.1045 & 1.16 &1.5& 1.26  &2.5& 2179 &  437 &  246&
                                                              1.30&2.9  & 2001  & 233 &  217&  0.02276\\
  E1024.10&0.0383& 1.06 &1.6& 1.09&2.3&  2194&  443 &  245&
                                                            1.10&2.6  &  2089  & 238&215 & 0.02566\\
\\
SDSS  &  & 2.5 && 3.5 &&1129 & 249 & 147&5.00   &&  916  &  154  &  140 & 0.01293 \\
SDSS &    &2.5 && 3.5 &&1129 & 249 & 147&5.40 &&    844 &   118 &   134 & 0.00981\\
\label{Tab1}                         
\end{tabular} 
\tablefoot{
The columns in the Table are as follows:\\ 
\noindent (1): sample name, where the last number shows the
redshift $z$;
\noindent (2):  $\sigma$ --  dispersion of the density contrast field;
\noindent (3): $P$ -- percolation  density threshold in mean density units;   
\noindent (4):  $x_P=(P-1)/\sigma$ -- reduced percolation  density threshold; 
\noindent (5): $D_{\mathrm{max}}$ --   density threshold at maxima of numbers of superclusters;
\noindent (6): $x_{\mathrm{max}}=(D_{\mathrm{max}}-1)/\sigma$ --
reduced density threshold at maxima of numbers of superclusters;
\noindent (7): $N_{\mathrm{max}}$ -- maximal number of superclusters;
\noindent (8):  $L_g$ -- geometrical diameter  of largest supercluster
in \Mpc\ at $D_{\mathrm{max}}$;
\noindent (9):  $L_f$ -- fitness diameter  of largest supercluster
in \Mpc\ at $D_{\mathrm{max}}$;
\noindent (10): $D_t$ -- density threshold to find
superclusters in mean density units; 
\noindent (11): $x_t = (D_t -1)/\sigma$ --  reduced density threshold to find
superclusters; 
\noindent (12): $N_{scl}$ -- number of superclusters at $D_t$;
\noindent (13):  $L_g$ -- geometrical diameter (length) of largest supercluster
in \Mpc\ at $D_t$;
\noindent (14):  $L_f$ -- fitness diameter (length) of largest supercluster
in \Mpc\ at $D_t$;
\noindent (15): $F_f$ --  total filling factor of over-density regions at
$D_t$.
}
\end{table*} 
}

Let us concentrate first to the behaviour of the model L1024 at the
present epoch, L1024.0.  At small threshold densities, $D_t \le 2$
($x \le 0$), there exists one percolating cluster, extending over the
whole volume of the computational box (here we use ``clusters'' as a
general term to designate over-density regions).  The percolation
threshold density, $P = D_t$, is defined as follows: for $D_t \le P$
there exists one and only one percolating cluster, for $D_t > P$ there
are no percolating clusters \citep{Stauffer:1979aa}.  Percolation
threshold densities, $P$, and reduced percolation threshold densities,
$x_P=(P-1)/\sigma$, are given in Table~\ref{Tab1}.  { As we see,
  the reduced  percolation threshold density of all models and epochs is 
  almost identical, $x_P \approx 1.5$.}
  In the reduced threshold density range $x \le 1.5$
geometrical diameters of clusters are equal to the diameter of the
box, $L_g = \sqrt{3}~L_0$, and their fitness diameters are equal to
the side-length of the box, $L_f=L_0$.

When we increase the threshold density, then at $x \approx 0$ there
appear additional clusters, and the number of clusters $N$ starts to
increase rapidly.  At percolating threshold, $x \approx 1.5$,
geometrical and fitness diameters of largest clusters, $L_g$ and
$L_f$, start to decrease: the large percolating cluster splits to
smaller clusters.  At $D_t =D_{\mathrm{max}} \approx 2.7$
($x_{\mathrm{max}} \approx 2.5$) the number of clusters reaches a
maximum, $N_{\mathrm{max}} \approx 8300$.  $D_{\mathrm{max}}$,
$x_{\mathrm{max}}$, $N_{\mathrm{max}}$ and respective geometrical and
fitness diameters of largest clusters at this threshold are given in
Table~\ref{Tab1}. At this threshold density clusters are still
complexes of large over-density regions, connected by filaments to
form systems of diameters $L_g \approx 300$~\Mpc\ and
$L_f \approx 200$~\Mpc, i.e. largest over-density regions are actually
complexes of superclusters.  The observed sample SDSS has similar
behaviour near $D_t= D_{\mathrm{max}}$.

When we increase the threshold density more, then the number of
clusters starts to decrease, since smallest clusters have maximal
densities lower than the threshold density, and disappear from the
cluster sample.
At $D_t \approx 4$ ($x \approx 4.5$) geometrical and fitness diameters of largest
clusters become close, $L_g \approx D_d \approx 150$~\Mpc.  With
further increase of the density threshold geometrical diameters
decrease, but fitness diameters have a minimum and thereafter start to
increase.  The reason for this behaviour is simple --- fitness
diameters are calculated from volumes of clusters by dividing
geometrical volumes to total filling factors,
$V_f(D_t) =V_{g}(D_t)/F_f(D_t)$.  At this threshold density range
the total filling factor of over-density regions, $F_f(D_t)$, decreases with
increasing $D_t$ more rapidly than the decrease of the geometrical
(i.e. the actual) volume of the largest clusters, $V_g(D_t)$.

An important aspect of this behaviour is the fact that fitness
diameters of largest clusters have a global minimum,
$L_f(D_t) \approx 140$ at $D_t =4.2$ ($x_t =5$ for the model L1024.0).
The geometrical diameter of largest clusters at this threshold density
is $L_g \approx 115$~\Mpc, similar to diameters of largest
superclusters known from catalogues by \citet{Einasto:2007tg,
  Liivamagi:2012}, based on 2dF and SDSS density fields.  This means,
that the global minimum of fitness diameters can be used as an
additional parameter to fix the threshold density to find
superclusters among clusters as supercluster candidates.  However,
here caution is needed.  In the model L1024.0 the region of low values
of the fitness diameters is rather large, and has local minima at
$x=2.8,~5.0,~8.0$.  Each of these minima marks breaks of the largest
cluster into smaller ones, see \citet{Liivamagi:2012}.

We denote the threshold density to find superclusters in our samples
as $D_t$ ($x_t$ in reduced threshold density units). { Threshold
  densities $D_t$ and $x_t$,  respective numbers
  of superclusters $N_{scl}$, geometrical and fitness lengths $L_g$ and
  $L_f$, are given in Table~\ref{Tab1}.  The mean reduced threshold
  density to find superclusters in our model samples has a large
  scatter with a mean value  $\approx 3.5$.}
At threshold density $D_t$ the total filling factor of high-density
regions lies in the interval $0.007 \ge F_f \ge 0.02$ (see
Table~\ref{Tab1}), and the respective correction factor to calculate
the fitness volumes has values $1/F_f \approx 100$. It is remarkable
that in spite of this large correction factor geometrical and fitness
diameters of largest superclusters are so similar.

\subsection{Changes of cluster diameters with time}

Supercluster geometrical diameter (length) functions of our model samples are
shown in Fig.~\ref{fig:evol} for redshifts $z=0$,
$z=1$, $z=3$, and $z=10$.  
At small threshold densities the over-density region extends over the
whole sample (largest clusters are percolated) and the geometrical
diameter of the largest cluster is equal to the diameter of the box.
With increasing threshold density the largest over-density region
splits into smaller units --- superclusters and their complexes ---
until only central regions of superclusters have densities higher than
the threshold density.  Geometrical diameters decrease with increasing
threshold density to a value about 30~\Mpc\ at $D_t=10$ ($x=14$ for the model
L1024.0).  This picture is shifted to lower threshold densities when
we consider earlier epochs at higher redshifts (diameters are
expressed in co-moving coordinates).  At epoch $z=10$ clusters exist
only at threshold densities $D_t \le 1.6$ ($x \le 9.5$).

\begin{figure*}[ht]
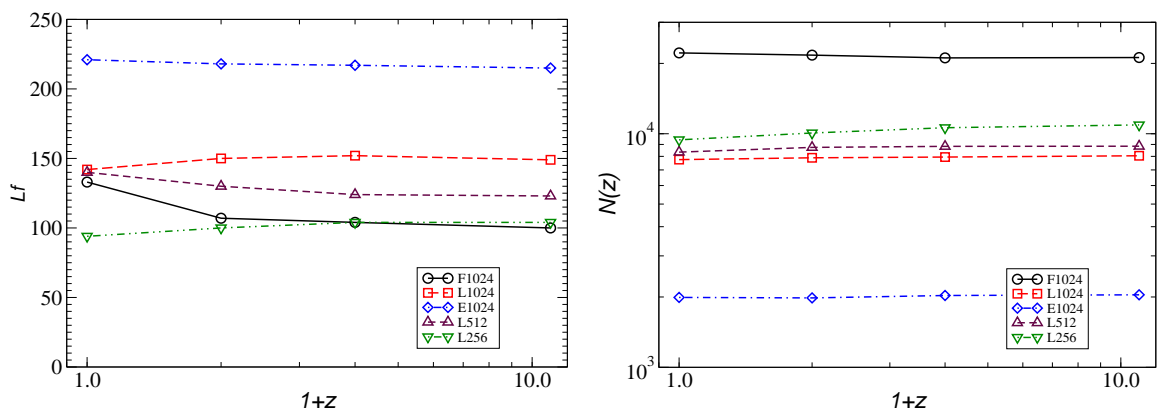
 
\centering 
\hspace{2mm} 
\resizebox{0.40\textwidth}{!}{\includegraphics*{evol_Lf-z.eps}}
\hspace{2mm} 
\resizebox{0.40\textwidth}{!}{\includegraphics*{evol_Nmax-z.eps}}\\
\hspace{2mm} 
\caption{{\em Left} panel shows the evolution of minimal fitness
  lengths with epoch, $L_f(z)$.  {\em Right} panel gives the evolution
  of the spatial density of maximal numbers of clusters with epoch,
  $N(z)$, per cubic cell of size 
  $L_0 =1$~\Gpc. Model designations as in Table~\ref{Tab1}. } 
\label{fig:evol4} 
\end{figure*}

The behaviour of fitness diameters is different --- they have a
minimum at a certain threshold density.  { Minimal fitness
  diameters of our models at various evolutionary epochs are given in
  Table~\ref{Tab1} and shown in Fig.~\ref{fig:evol4}}.  Minimal
fitness diameters of models are almost identical at all epochs (in
co-moving coordinates); for the model L1024 $L_f \approx
140$~\Mpc. Geometrical diameters at minima of fitness diameters are
$L_g \approx 115$~\Mpc\ for epochs $z \le 3$, and a bit more at $z=10$
(both in co-moving coordinates).

\subsection{Changes of cluster numbers  with time}

Right panels of Fig.~\ref{fig:evol} show numbers of clusters as
function of the reduced threshold density.  As noted above, at very
low threshold densities the whole over-density region contains one
percolating cluster since peaks of the density field are connected by
filaments to a connected region.  With increasing threshold density
some filaments became fainter than the threshold density, and the
connected region splits to smaller units. { At reduced threshold
  density $x \approx -0.5$ the number of clusters starts to increase
  rapidly with increasing threshold density.  The number of clusters
  reaches a maximum, $N_{\mathrm{max}}$, at threshold density
  $D_{\mathrm{max}}$.  The Table shows that at the earliest epoch the
  mean value of reduced threshold densities at maximum is
  $x_{\mathrm{max}} \approx 2.5$, increasing to
  $x_{\mathrm{max}} \approx 3.0$ at the present epoch; in the mean
  $x_{\mathrm{max}} = 2.9 \pm 0.5$.

  Fig.~\ref{fig:evol4} presents the evolution of maximal numbers of
  superclusters, $N(z)$; in this Figure numbers are actually spatial
  densities of superclusters, reduced to the volume of the sample of
  size $L_0 =1$~\Gpc.  Figs.~\ref{fig:evol}, \ref{fig:evol4} and
  Table~\ref{Tab1} show that maximal numbers of clusters are very
  similar at all evolutionary stages of the cosmic web,
  $N_{\mathrm{max}} \approx 8500$ for the model L1024. The almost
  constant reduced threshold density at maximum and the stability of
  the maximum itself are remarkable properties of the evolution of the
  cosmic web.  In most models the number of clusters at maxima is
  higher at earlier epochs, but only a bit.  This hints to the
  evolution: some small clusters have merged with larger clusters
  during the evolution.  However, the effect is surprisingly small. }

The decrease of the number of clusters with
increasing reduced threshold density $x$ after the maximum is more
rapid at earlier epochs.  At some threshold density highest peaks of
the density field are lower than the threshold density --- there are
no clusters at threshold higher than this limit.

\subsection{Influence of sample size}

To find the influence of sample size to the evolution of geometric
properties of superclusters in the cosmic web we used simulations in
boxes of sizes $L_0= 512$ and 256~\Mpc, with smoothing lengths
$R_B=8$~\Mpc.  Main results for both are given in Table~\ref{Tab1}
and  in Figs.~\ref{fig:evol}, \ref{fig:evol4}.  We see
that at all ages geometrical length functions of L512 models are
rather similar to respective functions of L1024 models.  Number
functions are also similar, but maximal numbers of clusters of the
L512 model are about 8 times lower than in the L1024 model, as
expected in a model having two times smaller box size. But spatial
densities of clusters are almost identical, see Fig.~\ref{fig:evol4}.

One difference of the L512 model from the L1024 one lies in the form
of the fitness length function: it has no well-defined global minimum.
There are four minima of lengths $L_f = 140 \pm 1$~\Mpc\ at threshold
densities $D_t = 3.2,~3.6~,5.0,~6.2$ { ($x = 3.4,~4.1,~6.2,~8.1$);
geometrical lengths at these threshold densities are
$L_g = 165,~155,~81,~39$~\Mpc, respectively. This shows that fitness
length minima alone are not sufficient to select superclusters: both
geometrical and fitness lengths are needed to have a proper choice. }

In the model L256 minima of fitness length functions are lower than in
models of larger box sizes, as seen in Table~\ref{Tab1} and
Figs.~\ref{fig:evol} and \ref{fig:evol4}.  Global minima of fitness
lengths are lower than in models of larger box sizes. Maximal numbers
of clusters are approximately 8 times lower than in the model L512,
but spatial densities of clusters are almost the identical. As in
models of larger box sizes maximal numbers of clusters at different
epochs are very close to each other, see Fig.~\ref{fig:evol4}.  The
scatter of all geometrical parameters is larger than in models of
larger box size, as expected.

The general behaviour of fitness length functions of L1024, L512 and
L256 models is also rather close.  Minima of fitness length functions at
different epochs have a spread $L_f =148 \pm 3$~\Mpc\ for the L1024
model, $L_f = 129 \pm 6$~\Mpc\ for the L512 model, and
$L_f = 100 \pm 5$~\Mpc\ for the L256 models.  This means that minima
of fitness functions are almost independent of the cosmic epoch, but
are smaller for models of smaller box sizes.  A likely explanation of this difference
is the size of models --- boxes of models L512 and L256 are not
large enough to fit very large density waves which are needed to form
largest superclusters. 

\begin{figure*}[ht]
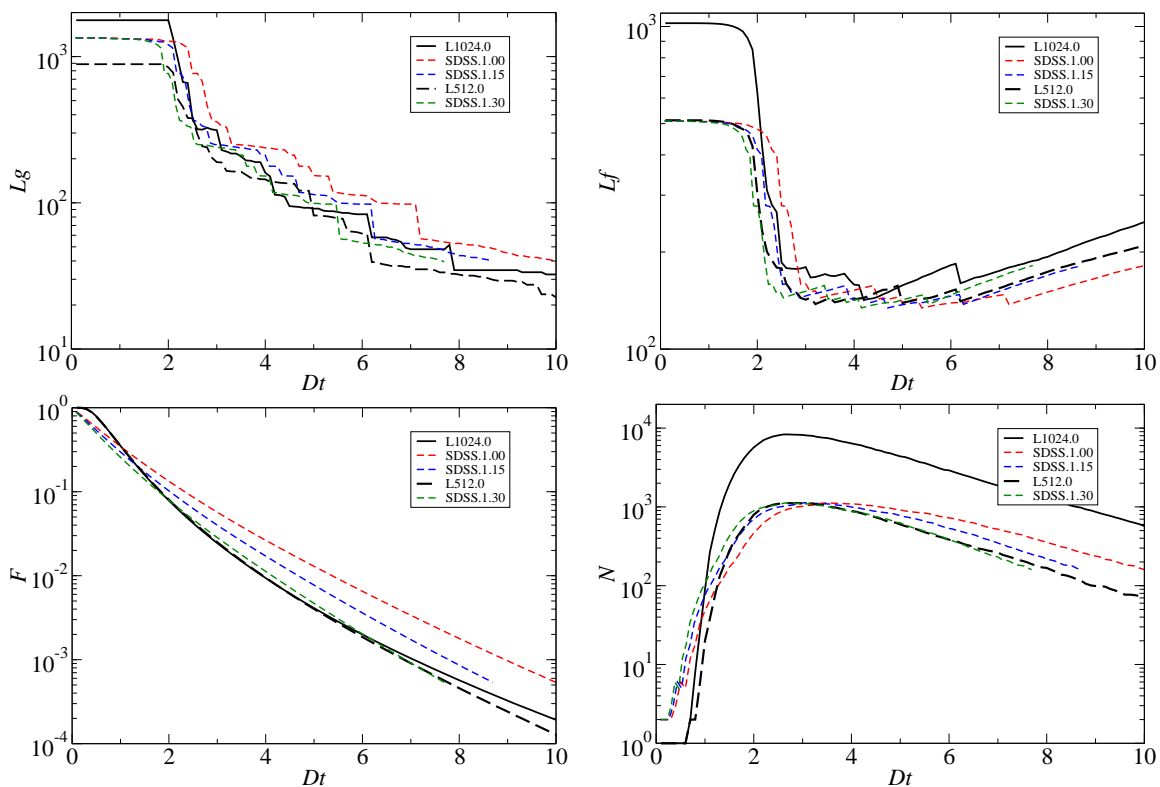
 
\centering 
\hspace{2mm}
\resizebox{0.40\textwidth}{!}{\includegraphics*{LCDM_Lgeom_D0trial3.eps}}
\hspace{2mm} 
\resizebox{0.40\textwidth}{!}{\includegraphics*{LCDM_Ldyn_D0trial3.eps}}\\
\hspace{2mm} 
\resizebox{0.40\textwidth}{!}{\includegraphics*{LCDM_ff_D0trial3.eps}}
\hspace{2mm} 
\resizebox{0.40\textwidth}{!}{\includegraphics*{LCDM_N_D0trial3.eps}}\\
\caption{The comparison of percolation functions of L1024.0 and L512.0
  models with SDSS samples. Model L1024.0 functions are plotted with
  bold lines, model L512.0 functions with bold dashed lines.
  Functions for SDSS samples are plotted with coloured dashed lines
  for biasing parameter values 1.00, 1.15, 1.30.  Top left panel of
  for geometrical length functions, top right panel for fitness length
  functions, bottom left panel for total filling factor functions,
  bottom right panel for number functions.}
\label{fig:sdss} 
\end{figure*} 

\begin{figure*}[ht]
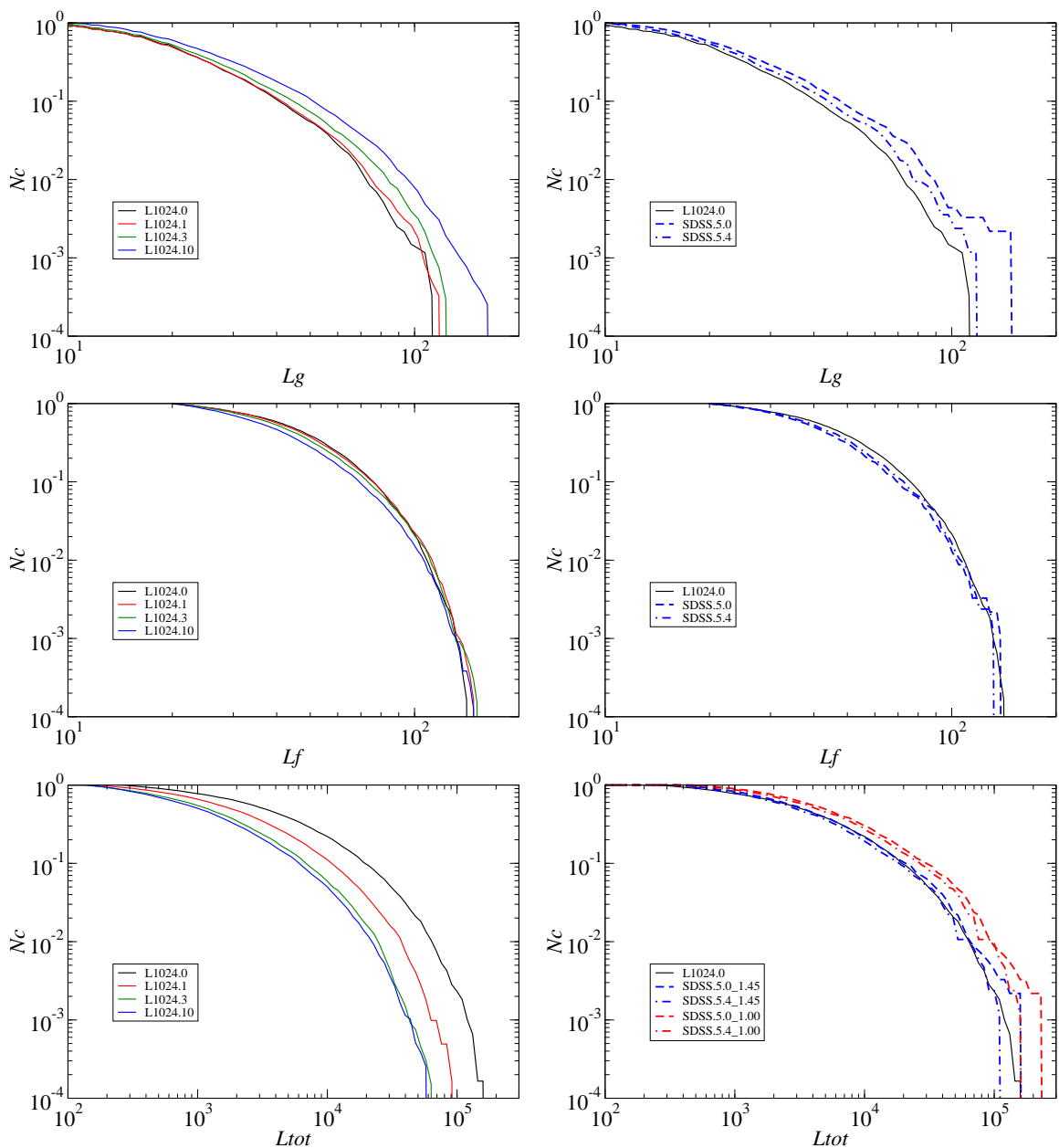
 
\centering 
\hspace{2mm} 
\resizebox{0.40\textwidth}{!}{\includegraphics*{LCDM_geomdiam-distr.eps}}
\hspace{2mm} 
\resizebox{0.40\textwidth}{!}{\includegraphics*{LCDM_SDSSgeomdiam-distr.eps}}\\
\hspace{2mm} 
\resizebox{0.40\textwidth}{!}{\includegraphics*{LCDM_dyndiam-distr.eps}}
\hspace{2mm} 
\resizebox{0.40\textwidth}{!}{\includegraphics*{LCDM_SDSSdyndiam-distr.eps}}\\
\hspace{2mm}  
\resizebox{0.40\textwidth}{!}{\includegraphics*{LCDM_lum-distr.eps}}
\hspace{2mm}  
\resizebox{0.40\textwidth}{!}{\includegraphics*{LCDM_SDSSlum-distr1.45.eps}}\\
\caption{{\em Left} panels show cumulative distribution of
  supercluster geometrical diameters, $L_g$, fitness diameters, $L_f$,
  and total luminosities, $\cal{L}$ of L1024 models at different
  evolution epochs, at {\em upper, central,  lower} panels, respectively.
  {\em Right} panels show the comparison of cumulative distributions
  of diameters and luminosities of L1024.0 model and SDSS samples.
  {\em Upper right}  panel shows the cumulative distributions of supercluster
  geometrical diameters, $L_g$, {\em central right} panel distributions of
  fitness diameters, $L_f$, {\em lower right} panel distributions of total
  luminosities, $\cal{L}$.  SDSS distributions are given for threshold
  densities $D_t =5.0,~5.4$; distributions of total luminosities are
  calculated for bias parameter $b=1.00$ (red) and $b=1.45$ (blue). }
\label{fig:diamdistr} 
\end{figure*}

\subsection{Influence of  smoothing length}

Superclusters have been traditionally searched using density fields
smoothed on 8~\Mpc\ scale.  To see how geometrical properties of
ensembles of clusters (over-density regions) depend on the smoothing
length we calculated percolation functions of the L1024 model using
smoothing lengths $R_B=4$~\Mpc\ and $R_B=16$~\Mpc; respective models
are designed as F1024 and E1024.
Percolation functions of these models are plotted in
Fig.~\ref{fig:evol}, main parameters of models
are given in Table~\ref{Tab1}.

In the model F1024 densities have a higher contrast than in the model
L1024.  The F1024 model selects smaller clusters (over-density
regions) than the L1024 model, thus maximal numbers of clusters are
about 3 times higher, see Fig.~\ref{fig:evol4}.  Global minima of
fitness lengths at different epochs are $L_f = 111 \pm 11$~\Mpc,
smaller than in the L1024 model, $L_f = 148 \pm 3$.

The model E1024 has lower density contrast than L1024 and F1024
models.  Global minima of fitness lengths of largest clusters are
larger than in models of the L1024 series,
$L_f \approx 218 \pm 2$~\Mpc.  Numbers of superclusters are about 4
times smaller than in models of the L1024 series, see
Fig.~\ref{fig:evol4}.  { Mean geometrical lengths of largest
  superclusters of the E1024 series are about two times larger than
  mean geometrical lengths of largest superclusters of the L1024
  series, see Table~\ref{Tab1}.}  The smoothing length $R_B=16$~\Mpc\
was used by \citet{Liivamagi:2012} to select superclusters from the
Luminous Red Giant (LRG) sample of the SDSS survey.  LRG
superclusters, found with the adaptive threshold density, are
approximately two times larger than superclusters of the SDSS main
galaxy sample.

Our analysis shows that smoothing scale is important in the selection
of supercluster type over-density regions.  Smaller smoothing selects
a larger number but smaller systems, and larger smoothing picks up
fewer number but larger systems.

\subsection{Comparison of model and SDSS supercluster ensembles}

In  Fig.~\ref{fig:sdss} we compare percolation functions of observed
SDSS samples with percolation functions of L1024.0 and L512.0 models at the
present epoch.  As we see, geometrical and fitness diameter functions
of SDSS samples are shifted relative to L1024.0 and L512.0 samples
towards higher threshold densities. The same effect is seen in filling
factor and number functions, presented in lower panels of
Fig.~\ref{fig:sdss}.  This is the well-known biasing effect.  All
densities are expressed in mean density units.  In model samples the
mean density includes, in addition to clustered matter, also dark matter in
low-density regions, where there are no galaxies, or galaxies are
fainter than the magnitude limit of the observational SDSS survey.  In
calculations of the mean density of the observed SDSS sample
unclustered and low-density dark matter is not included.  This means
that in the calculation of densities in mean density units densities
are divided to a smaller number, which increases density values of
SDSS samples \citep{Einasto:1999ku}.

{ We do not know how much matter is located in low-density regions
  with no galaxy formation.  Thus we estimated the biasing factor by
  an trial-and-error procedure.} 
We calculated corrected threshold densities
by dividing  threshold densities of SDSS samples by the density biasing factor, $b$:
\begin{equation}
(D_t)_c = D_t/b.
\end{equation} 
To select biasing factor values we tried a series of $b$ values 
$1.0 - 1.6$.  Percolation functions of SDSS samples are shown in
Fig.~\ref{fig:sdss} using three values of the density bias:
$b=1.00,~1.15,~1.30$.  The corrected supercluster diameter, filling
factor and number functions are in good agreement with L1024.0 and
L512.0 model functions using the biasing factor $b=1.30$.

\subsection{Distributions of diameters and luminosities}

In Fig.~\ref{fig:diamdistr} we show cumulative distributions of
geometrical and fitness diameters and luminosities of superclusters
for models of the L1024 series.  Data are given for all simulation
epochs, using threshold densities given in column (10) of
Table~\ref{Tab1}. 

As we see from the Fig.~\ref{fig:diamdistr},  geometrical diameters at
early epochs are larger than at the present epoch (in co-moving
coordinates), approximately by a factor of 2.  This means that in
co-moving coordinates superclusters shrink during the evolution.
Fitness diameters have a different behaviour --- the distribution of
fitness diameters is almost the same in co-moving coordinates at all
epochs.  This result means, that fitness diameters remain in co-moving
coordinates the same during 
the whole evolution of the cosmic web.
 
Cumulative distributions of geometrical and fitness diameters of SDSS
galaxies are shown in top right and middle right panels of
Fig. ~\ref{fig:diamdistr} for threshold densities $D_t=5.0,~5.4$.  We
see that the distribution of geometrical diameters is very sensitive
to the choice of the threshold density. The higher $D_t=5.4$ value is
suggested on the basis of the global minimum of fitness diameters.
This threshold density is also close to the threshold which yields
supercluster samples similar to \citet{Liivamagi:2012} supercluster
samples found with the adaptive threshold density.  For this threshold
density the largest SDSS supercluster has geometrical diameter,
$L_g = 118$~\Mpc, see Table~\ref{Tab1}.  The distribution found with
$D_t=5.0$ shifts the whole geometric diameter distribution towards
higher $L_g$ values.  Fitness diameter distributions of model and SDSS
samples are in good mutual agreement for both density threshold
values.

\begin{figure*}[ht]
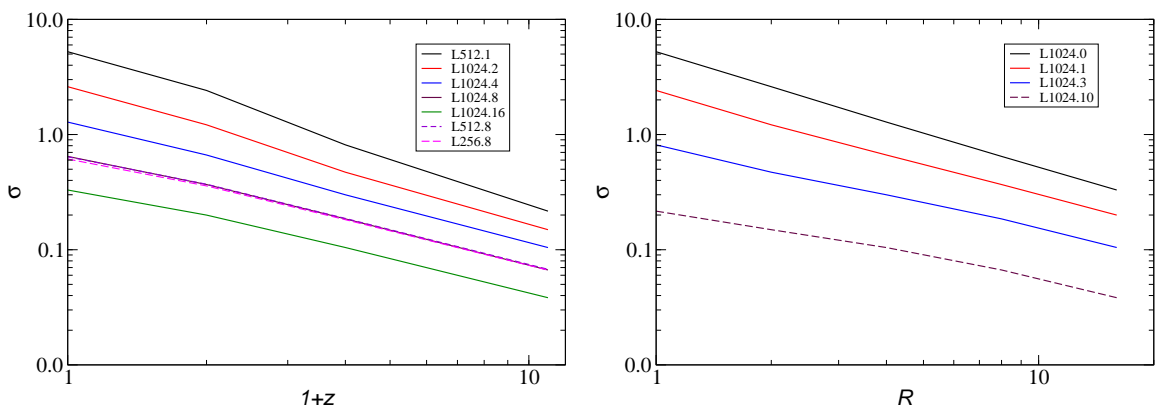
 
\centering 
\hspace{2mm}
\resizebox{0.40\textwidth}{!}{\includegraphics*{evol_ds-z.eps}}
\hspace{2mm}
\resizebox{0.40\textwidth}{!}{\includegraphics*{evol_s-R_B.eps}}\\
\caption{Left panel shows the change of the dispersion of density
  fluctuations $\sigma$ with cosmic epoch $z$ for our models.  In this
  panel we designated models as follows: L1024.i, L512.i, or L256.i,
  where $i=R_B$ is the smoothing kernel radius in \Mpc.  Right panel
  shows the dependence of $\sigma$ on density field smoothing length
  $R_B$; here index $i=z$ in model designation denotes the redshift
  $z$.  }
\label{fig:evol3} 
\end{figure*}

{ Lower left panel of Fig. ~\ref{fig:diamdistr} shows cumulative
  distributions of luminosities (actually masses) of L1024 model
  superclusters  at different epochs.  Luminosities are expressed in
  units of the mean mass of the model per cubic cell of size 1~\Mpc.} 
The comparison shows that masses 
of superclusters increase during the evolution, approximately by a
factor of three.  Early superclusters are less massive than at the
present epoch.  This result is in good agreement with simulations of
the growth of the cosmic web. The skeleton of the web with
superclusters forms already at early epoch.  Superclusters grow by the
infall of matter from low-density regions towards early forming knots
and filaments, forming early superclusters.

{ In lower right panel of Fig. ~\ref{fig:diamdistr} we compare
  cumulative distributions of luminosities of L1024 model and SDSS
  samples.  Luminosities of SDSS superclusters were calculated in
  units of mean luminosity densities in cells of size 1~\Mpc.  In this
  way model and observed distributions are comparable. To take into
  account the biasing effect in SDSS samples, we divided luminosities
  of SDSS superclusters to the biasing normalising factor
  $b=1.00,~1.45$.  As seen from the bottom right panel of
  Fig.~\ref{fig:diamdistr}, the correction $b=1.45$ brings total
  luminosity distributions of SDSS and L1024.0 samples to a very good
  agreement. This value of the correction factor is not far from the
  value, found above on the basis of percolation functions. }

We note that the number of L1024 model superclusters is approximately 8
times larger than the number of SDSS superclusters.  This difference
is expected due to the larger size of our model samples, 1024~\Mpc, about twice
the effective size of the SDSS main galaxy sample, 509~\Mpc.  In spite
of this difference in sample volume, diameter and luminosity
distributions of model and SDSS samples are very similar when proper
threshold densities and biasing corrections are applied.

\section{Discussion and summary}

{ \subsection{Dependence on the dispersion of the density contrast}

  The evolution of the cosmic web can be well described by percolation
  functions, using as argument the reduced threshold density,
  $x= (D_t - 1)/\sigma$, following \citet{Yess:1996aa, Sahni:1997ai,
    Colombi:2000fj}.  The dispersion (rms variance) of the density
  contrast, $\sigma$, was calculated using Eq.~(\ref{disp}) for all
  our models.  For completeness we calculated
  $\sigma$ also for models L1024 and L512 using smaller smoothing
  scales, $R_B = 1,~2$~\Mpc, as well as for other epochs, for which we
  had simulation output of density fields:
  $z=30,~10,~5,~3,~2,~1,~0.5,~0.0$.  The dispersion of the density
  contrast is a function of the cosmic epoch $z$ for constant
  smoothing scale, and of the smoothing scale $R_B$ for constant
  epoch.  Respective relations are shown in left and right panels of
  Fig.~\ref{fig:evol3}.  We see that there exists an almost linear
  relationship between $\sigma$ and $1+z$, and between $\sigma$ and
  $R_B$, when expressed in log-log format.  In spite of this
  similarity, ageing and smoothing affect the structure of the cosmic
  web in a very different way.  As expected, the parameter $\sigma$ is
  practically identical in models of various length $L_0$, when
  identical smoothing scale $R_B$ is applied.

  Now we consider the relationship between the dispersion of the
  density contrast $\sigma$ and the percolation threshold density,
  $P$.  Data given in Table~\ref{Tab1} show that there exists an
  almost linear relationship between $\sigma$ and percolation
  threshold $P$. Most importantly,  all our models of different
  length $L_0$ and smoothing scale $R_B$ lie close to an identical
  curve, which can be written as follows: $P = 1+ 1.5\times \sigma$.
  This relationship is expected since in the very early universe when
  $\sigma \rightarrow 0$ the percolation threshold density approaches
  $P \rightarrow 1$ \citep{Einasto:2018aa}.  Reduced percolation
  threshold densities $x_P =(P-1)/\sigma$ are given in
  Table~\ref{Tab1}.  The mean value for our five models is
  $x_P = 1.49 \pm 0.13$, in good agreement with results by
  \citet{Colombi:2000fj}.

  A similar relationship exists also for density thresholds, corresponding to
  maxima of numbers of superclusters,
  $D_{\mathrm{max}} = 1+ 2.9\times \sigma$.  Reduced density thresholds
  at maxima of numbers of superclusters,
  $x_{\mathrm{max}} = (D_{\mathrm{max}} -1)/\sigma$, are given in
  Table~\ref{Tab1}.  As noted above at the earliest epoch the mean value is
  $x_{\mathrm{max}} \approx 2.5$, increasing to
  $x_{\mathrm{max}} \approx 3.0$ at the present epoch. }

\subsection{Fitness diameters as parameters of the cosmic web}

{ Fitness volumes (and respective diameters) are geometrical
  parameters, proportional to the volume of the largest supercluster,
  divided to the volume of all over-density regions at the given
  threshold density.  Fitness volumes of largest clusters are
  approximately inversely proportional to the number of clusters.  But
  fitness volumes and numbers of clusters are calculated from
  different data, from volumes of largest superclusters and total
  number of clusters, respectively.  Thus these parameters represent
  different aspects of the structure of the cosmic web.

\begin{figure*}[ht] 
\centering 
\resizebox{0.95\textwidth}{!}{\includegraphics*{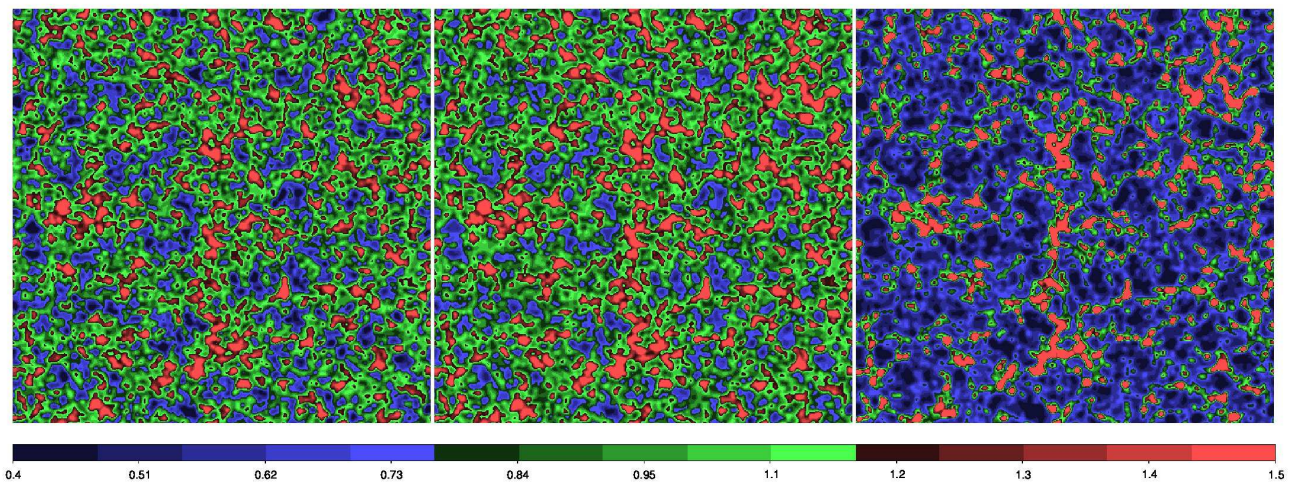}}\\
\caption{Density fields of the L1024.10, L1024.3 and L1024.0 models,
  found with smoothing kernel of radius 8~\Mpc.  {\em Left} panel
  corresponds to the epoch $z=10$, {\em middle} panel to epoch $z=3$,
  {\em right} panel to the present epoch $z=0$.  Cross-sections are
  shown in a 2~\Mpc\ thick layer, densities are expressed in linear
  scale. Colour scales from left to right are:
  $0.8 - 1.2,~ 0.4 - 1.5, 0.2 - 2.5$. }
\label{fig:Mdenfield} 
\end{figure*}

An essential property of the fitness diameter functions is the
presence of global minima at certain threshold densities.  The fitness
diameter function has a number of local minima, showing 
the presence of  breaks, where largest superclusters
split to smaller units.  For a detailed discussion of this phenomenon
see \citet{Liivamagi:2012}.  
Breaks of fitness length functions (and
breaks of geometrical length functions) are different
in models of different size, smoothing scale and epoch, and have a
rather large scatter.  To select the proper value of the threshold
density to find superclusters we used local minima of the fitness
length function, which correspond to geometrical lengths of largest
superclusters, close to lengths, usually accepted for largest SDSS
superclusters \citep{Einasto:2007tg, Costa-Duarte:2011ys,
  Luparello:2011fr, Liivamagi:2012}.  For the mean value of the
reduced density threshold to select superclusters we get 
$x_t = 3.44 \pm 0.76$.

Fitness diameters of superclusters near minima are approximately
identical in samples of different size.  Largest
superclusters in samples of smaller size are only slightly smaller
than largest superclusters in samples of larger size.  Cluster
numbers  are approximately proportional to the volume of the sample,
thus cluster numbers, reduced to identical sample volume are very
close, see right panel of Fig.~\ref{fig:evol4}.  }  The dependence
of fitness diameters on the sample size up to $L_0=1024$~\Mpc\
suggests that samples smaller than this scale do not represent fair
samples of the Universe for the formation of representative samples of
rich superclusters.  On the other hand, scales larger than $\sim
1000$~\Mpc\ have little effect on the structure of the cosmic web, as
suggested by \citet{Klypin:2018aa}. Thus we can take our models of the L1024 series
as estimates of fair samples of the Universe.  This scale is larger
than expected from previous analyses \citep{Einasto:1993ac}.

{ Fitness diameters of largest superclusters depend on the smoothing
scale used to select superclusters.  This property is expected, since 
smoothing highlights properties of the cosmic web on different
scales.  When one uses very small smoothing of the order of 1~\Mpc,
one gets as characteristic elements of the web  giant galaxies of the
M31 and Milky Way type, surrounded by dwarf satellites, as well as
small groups and clusters of galaxies. Smoothing with scale 4~\Mpc\
highlights systems  of intermediate scale between clusters
and traditional superclusters.  Smoothing with 8~\Mpc\ scale selects
ordinary  superclusters.  Smoothing with 16~\Mpc\
scale corresponds to rich superclusters, selected on the basis of bright
LRG galaxies, as done by \citet{Liivamagi:2012}.  }

\subsection{Evolution of the ensemble of superclusters}

{ An important aspect of percolation functions is their shape.
  Fig.~\ref{fig:evol} shows that the shape of percolation functions is
  almost identical for all models and epochs for $x \le 1.5$.  The
  shape of fitness length and number functions is approximately
  symmetrical around the value $x=x_{\mathrm{max}} \approx 2.5$ at
  early epoch $z=10$.  This means that at these scales the growth of
  density perturbations is nearly linear.  At later epochs the maximum
  of number functions is shifted to $x=x_{\mathrm{max}} \approx 3.0$.
  As the web evolves, fitness length and number functions are
  gradually shifted towards higher $x$-values and the symmetry is
  gradually lost.   In the model F1024 with smaller smoothing scale
  the asymmetry growth is the largest.  }

Supercluster luminosity functions (distributions of luminosities of
superclusters) of L1024.0 model and SDSS samples are very similar when
a biasing correction is taken into account.  Model and SDSS luminosity functions
are rather close to luminosity functions found by
\citet{Einasto:2006kl} for early SDSS and Two-degree-Field (2dF)
superclusters.  It is unclear why model superclusters found by
\citet{Einasto:2006kl} on the basis of Millennium simulations
\citep{Croton:2006qy}, had a different luminosity function.  In this
paper we used identical procedures to select superclusters based on
density fields smoothed with 8~\Mpc\ kernel, thus present results
should be more reliable.

Arguments based on geometrical and fitness diameter functions suggest
that very large over-density regions, such as the Sloan Great Wall and
the BOSS Great Wall, are actually complexes of superclusters, as
studied by \citet{Liivamagi:2012, Einasto:2016ve, Einasto:2017kx}.
Similarly the Laniakea Supercluster, introduced by \citet{Tully:2014},
is a complex of several previously known superclusters: the Local
Supercluster, the Great Attractor, and some smaller cluster filaments and
clouds. The Laniakea Supercluster is surrounded by rich Coma,
Perseus-Pisces, Hercules and Shapley Superclusters.

\subsection{Cocoons of the cosmic web}

{ To understand better the evolution of the cosmic web on
  supercluster scale, we show in Fig.~\ref{fig:Mdenfield} the visual
  appearance of density fields of models L1024 at different epochs: in
  the left panel at the early epoch $z=10$, in the middle panel at
  epoch $z=3$, and in the right panel at the present epoch $z=0$, all
  smoothed with 8~\Mpc\ co-moving scale. The evolution of density
  fields can be followed by comparison of panels.  This comparison
  suggests that supercluster-type structural elements of the
  cosmic web are present already at very early epochs.  Of course,
  there are differences on small scales, but main supercluster-type
  elements of the web are seen at similar locations at all
  epochs. Basic visible changes are the increase of the density
  contrast: distributions of densities at epochs $z=10$ and $z=3$ are
  very similar, only the amplitude of density perturbations has
  increased.  This means that in this time interval the evolution is near
  to a linear growth.  On later epochs the non-linearity of the
  evolution is dominant.  The flow of small-scale structural elements
  towards large ones is more visible.

  Elements of the cosmic web evolve with time.  Physical clusters of
  galaxies grow by merging of smaller clusters and by infall of
  non-clustered matter, filaments merge, and voids became 
  emptier.  Superclusters also change, their sizes shrink in co-moving
  coordinates, and masses grow by infall and merging.  Similar general
  visual appearance of the density fields at very early and present
  epochs suggests that supercluster embryos were created very
  early. This result is not surprising, already \citet{Kofman:1988}
  demonstrated that the whole present-day structure is seen in the
  initial fluctuation distribution.

  \citet{Tully:2014} defined superclusters as ``basins of
  attraction'': supercluster is the volume containing all galaxies and
  particles whose flow lines converge at a given attractor, the local
  minimum of the gravitational potential.

  We prefer to define superclusters as high-density regions of the
  cosmic web. Tully et al. ``basins of attraction'' are in our
  terminology supercluster cells of dynamical influence, for short we
  can call these cells cocoons.  Cells of dynamical influence are
  regions around superclusters, from which superclusters collect their
  matter.  They are separated from each other by surfaces, where on
  the one side the smoothed velocity flow is directed to one
  supercluster, and on the other side to an another supercluster.  In
  this way the whole volume of the universe is divided into
  supercluster cells of dynamical influence.  Cells of dynamical
  influence are different from cells introduced by
  \citet{Joeveer:1977lj, Joeveer:1978pb} (see also
  \citet{Aragon-Calvo:2010}), which are cellular low-density regions
  surrounded by a network of high-density structures: clusters,
  filaments and walls.

  Our analysis gives support to the presence of supercluster cells.
   Main arguments are the following: (i) almost constant
  number of superclusters and approximately constant fitness diameters
  in co-moving coordinates at different cosmical epochs; (ii) growth
  of the mass of superclusters and decrease of supercluster geometric
  diameters (in co-moving coordinates) with time; (iii) visual
  appearance of density fields of models at various evolutionary
  epochs, smoothed with co-moving scale 8~\Mpc.

  Supercluster cocoons are seen in all our models using different box
  sizes, and their presence is an important property of the cosmic
  web.  This suggests that the essential evolution of superclusters
  occurs inside supercluster cocoons.  Supercluster cocoons have
  volumes about hundred times larger than geometrical volumes of
  superclusters.  Fitness diameters of largest superclusters depend
  slightly on the size of the model and on the smoothing length used
  in calculation of the density field.  Smoothing highlights
  properties of the cosmic web at various scales.  Thus the size of
  supercluster cocoons is not a physical scale as the Baryonic
  Acoustic Oscillation (BAO) scale.  BAO phenomenon is caused by
  baryonic oscillations of hot gas before the cosmic recombination.
  Seeds of the cosmic web are scale-free primordial fluctuations.  The
  cosmic web has a fractal nature, and superclusters are elements of
  the cosmic web which can be highlighted by smoothing. }

\subsection{Summary remarks}

We investigated evolutionary changes of geometrical properties of the
conventional $\Lambda$CDM model applying an extended percolation
analysis, which characterises general geometrical properties of the
ensemble of superclusters.  We calculated density fields of the
$\Lambda$CDM model using three sample box sizes
$L_0=1024,~512,~256$~\Mpc, and made the analysis for four evolutionary
epochs of the Universe: $z=0,~1,~3,~10$. The analysis was made using
density fields smoothed with an $R_B=8$~\Mpc\ kernel; for comparison
also smoothing with 4 and 16~\Mpc\ kernels was done.  We scanned
density fields in a wide interval, and found connected over-density
regions (clusters).  Lengths, total filling factors, and
numbers of largest clusters as functions of the threshold density were
used as percolation functions.  {  In the analysis we used threshold
densities in units of the mean density of the sample,
$D_t$, and reduced threshold densities, $x = (D_t - 1)/\sigma$, were
$\sigma$ is the dispersion of the density contrast, $D-1$. }
In addition to geometrical diameters
we used fitness diameters, calculated on the basis of cluster volumes
and total filling factors.

Our basic methodical contribution to the percolation analysis is the
addition of fitness volumes and diameters of clusters (superclusters)
to the list of geometrical properties. We found that the fitness
diameter of superclusters is a stable parameter, useful to
characterise sizes of superclusters, and to study geometrical
properties of the cosmic web.  Fitness diameters of superclusters as
functions of the threshold density have a global minimum.  { Near
  the minimum of  fitness diameters numbers of superclusters have a
  maximum.  At this density threshold the  cosmic
  web can be divided into supercluster cells. }

The basic conclusions of our study are as follows.
 
\begin{enumerate}

\item{} Minimal fitness diameters of largest superclusters almost do not
  change during the evolution of the cosmic web (in co-moving
  coordinates). 

\item{} Numbers of superclusters as a function of the threshold
  density have maxima which are approximately
  constant for all evolutionary epochs.
  
\item{} { The maximum of supercluster numbers and minimum of
    fitness diameters occurs in all models at reduced threshold
    density, $x_{\mathrm{max}} \approx 2.5$ at early evolutionary
    epoch, increasing to $x_{\mathrm{max}} \approx 3.0$ at the present
    epoch.
  
  \item{} The shape of percolation functions is very similar in models
    of various age and smoothing scale.  At early epoch percolation
    functions around $x_{\mathrm{max}}$ are approximately symmetrical,
    showing nearly linear growth of density perturbations.  At later
    epochs the positive wing of fitness length and number functions
    increases, showing the growing non-linearity of density
    perturbations. }

\item{} Geometrical diameters of superclusters decrease during the
  evolution (in co-moving coordinates); luminosities of superclusters
  increase during the evolution.

\item{} { Essential evolutionary changes occur inside supercluster
    cells or cocoons. Volumes of supercluster cells are about hundred
    times larger than their geometrical volumes.}

\end{enumerate}

In the present study we used data on spatial coordinates which allowed
to test the concept of supercluster cells as representatives of true
dynamical volumes.  Our study confirms that the concept of
supercluster cells (basins of attraction) has cosmological significance.  The
determination of true dynamical volumes using velocity data and the
gravitation potential field would be an interesting task.

Our study also showed that percolation functions of model samples
deviate in a very clear way from respective observed functions derived
using SDSS galaxy samples.  Differences can be understood in terms of
the biased galaxy formation, where in low-density regions galaxies do not
form, or are too faint to fall into the magnitude range covered by
SDSS observations.  A more detailed investigation of the biasing
phenomenon using density fields of models and galaxies is an
interesting task, but outside the scope of the present study.

\begin{acknowledgements} 

  {  Authors thank the anonymous referee for stimulating suggestions.  }
  This work was supported by institutional research funding IUT26-2
  and IUT40-2 of the Estonian Ministry of Education and Research. We
  acknowledge the support by the Centre of Excellence``Dark side of
  the Universe'' (TK133) financed by the European Union through the
  European Regional Development Fund.  The study has also been
  supported by ICRAnet through a professorship for Jaan Einasto, and
  by the University of Valencia (Vicerrectorado de Investigaci\'on)
  through a visiting professorship for Enn Saar and by the Spanish MEC
  projects ``ALHAMBRA'' (AYA2006-14056) and ``PAU'' (CSD2007-00060),
  including FEDER contributions.

  We thank the SDSS Team for the publicly available data releases. 
  Funding for the SDSS and SDSS-II has been provided by the Alfred 
  P. Sloan Foundation, the Participating Institutions, the National 
  Science Foundation, the U.S. Department of Energy, the National 
  Aeronautics and Space Administration, the Japanese Monbukagakusho, 
  the Max Planck Society, and the Higher Education Funding Council for 
  England. The SDSS Web Site is \texttt{http://www.sdss.org/}.

The SDSS is managed by the Astrophysical Research Consortium for the 
Participating Institutions. The Participating Institutions are the 
American Museum of Natural History, Astrophysical Institute Potsdam, 
University of Basel, University of Cambridge, Case Western Reserve 
University, University of Chicago, Drexel University, Fermilab, the 
Institute for Advanced Study, the Japan Participation Group, Johns 
Hopkins University, the Joint Institute for Nuclear Astrophysics, the 
Kavli Institute for Particle Astrophysics and Cosmology, the Korean 
Scientist Group, the Chinese Academy of Sciences (LAMOST), Los Alamos 
National Laboratory, the Max-Planck-Institute for Astronomy (MPIA), 
the Max-Planck-Institute for Astrophysics (MPA), New Mexico State 
University, Ohio State University, University of Pittsburgh, 
University of Portsmouth, Princeton University, the United States 
Naval Observatory, and the University of Washington.

\end{acknowledgements}

\bibliographystyle{aa} 

\end{document}